\documentclass[aps,preprint]{revtex4}%
\usepackage{amsfonts}
\usepackage{amsmath}
\usepackage{amssymb}
\usepackage{graphicx}%
\providecommand{\U}[1]{\protect\rule{.1in}{.1in}}

\begin{document}
\title[ ]{The Planck Blackbody Spectrum Follows from the Structure of Relativistic Spacetime}
\author{Timothy H. Boyer}
\affiliation{Department of Physics, City College of the City University of New York, New
York, New York 10031}

\begin{abstract}
Here we show that within classical physics, the Planck blackbody spectrum can
be derived directly from the structure of relativistic spacetime. \ In
noninertial frames, thermal radiation at positive temperature is connected
directly to zero-point radiation whose spectrum follows from the geodesic
structure of the spacetime. \ The connection between zero-point radiation and
thermal radiation at postive temperature is through a time-dilating conformal
transformation in the noninertial frame. \ Transferring the spectrum back to
Minskowski spacetime, the Planck spectrum is obtained.

\end{abstract}
\maketitle

In textbooks of physics, the Planck blackbody radiation spectrum is said to
require quantum theory for its derivation. \ Actually, the Planck blackbody
spectrum can be derived directly from the structure of relativistic spacetime
within classical physics. \ In noninertial frames, thermal radiation at
positive temperature is connected directly to zero-point radiation whose
spectrum follows from the geodesic structure of the spacetime.

The derivation of the Planck spectrum for relativistic radiation depends upon
the following essential ideas. \ Zero-point radiation corresponds to the
spectrum of random classical radiation which is as featureless as possible;
its correlation function depends upon only the geodesic separation between the
points at which the correlation function is evaluated. \ A Minkowski frame in
flat spacetime is such a specialized, featureless system that Minkowski
spacetime gives no structure to the zero-point spectrum; in a Minkowski frame,
zero-point radiation is carried into itself by a time-dilating symmetry
transformation. \ However, in a static noninertial frame, each spatial point
undergoes a proper acceleration relative to the momentarily comoving reference
frame, and this acceleration gives a structure to the zero-point correlation
function; under a time-dilating symmetry transformation, zero-point radiation
is carried into thermal radiation at positive temperature. \ The thermal
radiation spectrum in the noninertial frame can then be carried back into an
inertial frame. \ The result is the Planck spectrum for blackbody radiation in
a Minkowski frame derived from fundamental ideas of spacetime structure. \ In
this article, we carry out the explicit calculations for relativistic scalar
radiation in a spacetime of four dimensions. \ 

In a Minkowski frame in flat spacetime, the familiar spacetime coordinates
$ct,x,y,z$ are geodesic coordinates with interval $ds^{2}=g_{\text{M}\mu\nu
}dx^{\mu}dx^{\nu}=c^{2}dt^{2}-dx^{2}-dy^{2}-dz^{2}.$ \ Thus the geodesic
separation between two spacetime points (primed and unprimed) is simply
$s^{2}=g_{\text{M}\mu\nu}(x^{\mu}-x^{^{\prime}\mu})(x^{\nu}-x^{^{\prime}\nu
})=c^{2}(t-t^{\prime})^{2}-(x-x^{\prime})^{2}-(y-y^{\prime})^{2}-(z-z^{\prime
})^{2}.$ \ The relativistic scalar field $\phi$ with Lagrangian density
$\mathcal{L}=[1/(8\pi)]g^{\mu\nu(}\partial\phi/\partial x^{\mu})(\partial
\phi/\partial x^{\nu})$ has scaling dimension one in four spacetime
dimensions. \ Accordingly, the correlation function for the zero-point
radiation of the relativistic scalar field must take the form%

\begin{equation}
\left\langle \phi_{0}(ct,x,y,z)\phi_{0}(ct^{\prime},x^{\prime},y^{\prime
},z^{\prime})\right\rangle =\frac{-(\hbar c/\pi)}{c^{2}(t-t^{\prime}%
)^{2}-(x-x^{\prime})^{2}-(y-y^{\prime})^{2}-(z-z^{\prime})^{2}}\label{zcorr}%
\end{equation}
where the constant in the numerator has been chosen so as to connect with the
familiar spectrum of classical zero-point radiation corresponding to an energy
$\mathcal{E}_{0}(\omega)=(1/2)\hbar\omega$ per normal mode.\cite{any}%
\cite{rel} \ We notice that the correlation function depends simply upon the
geodesic separation between the spacetime points, and there is no
characteristic time or length in this correlation function; the one constant
$(\hbar c/\pi)$ which appears gives an overall scale. \ The appearance of the
constant $\hbar$ has nothing to do with quantum theory and is merely a
constant taking the numerical value of Planck's constant so as to give
agreement with experimental measurements of Casimir forces.\cite{any}

The correlation function for zero-point radiation in Eq. (\ref{zcorr}) is
invariant under a time-dilating symmetry transformation. \ The Minkowski
spacetime metric undergoes a scale transformation under the time-dilating
transformation $t\rightarrow\overline{t}=\sigma t$ where $\sigma$ is a
positive numerical constant, provided that the space coordinates are also
dilated giving the full transformation
\begin{equation}
t\rightarrow\overline{t}=\sigma t,~x\rightarrow\overline{x}=\sigma
x,~y\rightarrow\overline{y}=\sigma y,~z\rightarrow\overline{z}=\sigma
z,\label{dil}%
\end{equation}
so that the spacetime interval is rescaled as
\begin{equation}
d\overline{s}^{2}=\sigma^{2}ds^{2}=\sigma^{2}(c^{2}dt^{2}-dx^{2}-dy^{2}%
-dz^{2}).\label{mdil}%
\end{equation}
$~$\ If we interpret this time-dilating symmetry transformation in the active
sense as a transformation of the field $\phi$ in the old coordinates, then we
can write the transformation as $\phi\rightarrow\overline{\phi}$ where
$\overline{\phi}(ct,x,y,z)=\sigma\phi(c\sigma t,\sigma x,\sigma y,\sigma z).$
\ The correlation function for zero-point radiation $\phi_{0}$ is invariant
under this transformation\cite{conf} since
\begin{align}
\left\langle \overline{\phi}_{0}(ct,x,y,z)\overline{\phi}_{0}(ct^{\prime
},x^{\prime},y^{\prime},z^{\prime})\right\rangle  &  =\left\langle \sigma
\phi_{0}(c\sigma t,\sigma x,y,\sigma z)\sigma\phi_{0}(c\sigma t^{\prime
},\sigma x^{\prime},\sigma y^{\prime},\sigma z^{\prime})\right\rangle
\nonumber\\
&  =\frac{-\sigma^{2}(\hbar c/\pi)}{c^{2}(\sigma t-\sigma t^{\prime}%
)^{2}-(\sigma x-\sigma x^{\prime})^{2}-(\sigma y-\sigma y^{\prime}%
)^{2}-(\sigma z-\sigma z^{\prime})^{2}}\nonumber\\
&  =\frac{-(\hbar c/\pi)}{c^{2}(t-t^{\prime})^{2}-(x-x^{\prime})^{2}%
-(y-y^{\prime})^{2}-(z-z^{\prime})^{2}}.\label{invar}%
\end{align}

In contrast to zero-point radiation, equilibrium thermal radiation at positive
temperature in a Minkowski frame will involve a parameter corresponding to the
temperature $T$ in the unique inertial frame in which the radiation spectrum
is isotropic. Now within classical physics, the zero-point radiation
($\mathcal{E}_{0}(\omega)=(1/2)\hbar\omega$ per normal mode) is regarded as
radiation which is always present;\ thermal radiation ($\mathcal{E}_{T}%
(\omega,T)$ per normal mode), on the other hand, involves additional radiation
energy per normal mode above the zero-point energy per normal mode,
$\mathcal{E}_{T}>\mathcal{E}_{0}$. \ Since thermal radiation involves\ a
finite total radiation energy density above the zero-point radiation for a
system of an infinitely many normal modes, the smooth thermal spectrum must
involve a transition frequency $\omega_{tr}$ where the zero-point radiation
energy per normal mode (which vanishes at low frequency and increases
steadily) is equal to the thermal energy per normal mode of the lowest
frequency modes, $\mathcal{E}_{T}(0,T)=k_{B}T=\mathcal{E}_{0}(\omega
_{tr})=(1/2)\hbar\omega_{tr};$ therefore $\omega_{tr}=2k_{B}T/\hbar.$ This
transition frequency corresponds roughly to the frequency in the spectrum
where the functional form changes from the low-frequency to the high-frequency
form. \ The correlation function for thermal radiation will contain a
transition time $t_{tr}=2\pi/\omega_{tr}$ which reflects this transition
frequency. \ Under the time-dilating symmetry which rescales the radiation
fields as $\overline{\phi}(ct,x,y,z)=\sigma\phi(c\sigma t,\sigma x,\sigma
y,\sigma z)$, the thermal spectrum will be transformed from positive
temperature $T$ over to positive temperature $\overline{T}=\sigma T,$ and the
transition frequency $\omega_{tr}$ and transition time $t_{tr}$ will be
transformed to $\overline{\omega}_{tr}=\sigma\omega_{tr}$ and $\overline
{t}_{tr}=t_{tr}/\sigma.$ \ We notice that for any positive dilation factor
$\sigma,$ the new temperature $\overline{T}$, the new transition frequency
$\overline{\omega}_{tr},$ and the new transition time $t_{tr}$\ are always
positive. \ Thus under a time-dilating symmetry transformation, the set of
thermal correlation functions and the set of thermal spectral functions for
finite temperature in a Minkowski frame are carried into themselves, but the
zero-point correlation function and zero-point radiation spectrum are not
included in these sets. \ The zero-point correlation function and zero-point
spectrum are the limiting functions obtained from the thermal radiation
functions at positive temperature as $\sigma\rightarrow0.$ \ Crucially, we
cannot go back from this $\sigma\rightarrow0$-limit to obtain thermal
radiation at positive temperature from zero-point radiation at zero
temperature in a Minkowski frame. \ The Minkowski frame is such a specialized
system that its coordinates are geodesic coordinates, and thermal radiation
cannot be obtained from zero-point radiation by a time-dilating symmetry transformation.

The situation is quite different for a static coordinate frame in a
gravitational field or for a static noninertial coordinate frame in flat
spacetime. \ In a static coordinate frame, none of the metric components
$g_{\mu\nu}$\ are functions of the time coordinate. \ Although relativistic
radiation follows the light-like geodesics in a relativistic spacetime, the
points with fixed spatial coordinates in a noninertial frame do not follow
geodesics. \ In a static coordinate frame with time coordinate $\eta$, we have
$ds^{2}=g_{00}(x^{i})d\eta^{2}+g_{ij}(dx^{j})(dx^{i})$ with $\partial
_{0}g_{\mu\nu}=\partial_{\eta}g_{\mu\nu}=0$ where the indices $i$ and $j$
label the spatial coordinates. Thus in a static noninertial frame, the
geodesic equation $d^{2}x^{\mu}/d\tau^{2}+\Gamma_{\rho\sigma}^{\mu}(dx^{\rho
}/d\tau)(dx^{\sigma}/d\tau)=0$ for a particle instantaneously at rest
$(dx^{i}/d\eta)=0$ at spatial coordinates $x^{i}$ becomes\cite{Carroll153}%
\begin{equation}
0=\frac{d^{2}x^{\mu}}{d\tau^{2}}+\Gamma_{00}^{\mu}\left(  \frac{d\eta}{d\tau
}\right)  ^{2}=\frac{d^{2}x^{\mu}}{d\tau^{2}}+\left(  -\frac{1}{2}%
g^{\mu\lambda}\partial_{\lambda}g_{00}\right)  \frac{c^{2}}{g_{00}%
}.\label{geod}%
\end{equation}
Thus a point with fixed spatial coordinates $x^{i}$ in the static coordinate
frame has a constant proper acceleration given by $-(1/2)g^{\mu\lambda
}(\partial_{\lambda}g_{00})c^{2}/g_{00},$ which is the negative of the
geodesic acceleration in Eq. (\ref{geod}). \ 

A Rindler frame in flat spacetime is an example of a simple static noninertial
frame. \ In a Rindler frame where
\begin{equation}
ds^{2}=\xi^{2}d\eta^{2}-d\xi^{2}-dy^{2}-dz^{2},\label{gRind}%
\end{equation}
the proper acceleration $a(\xi)$ of a point with coordinates $x^{1}=\xi
,~x^{2}=y,~x^{3}=z$ is in the direction of the coordinate $x^{1}=\xi,$ and is
given by
\begin{equation}
a(\xi)=-\frac{1}{2}g^{11}(\partial_{1}g_{00})c^{2}/g_{00}=-\frac{1}%
{2}(-1)(2\xi)\frac{c^{2}}{\xi^{2}}=\frac{c^{2}}{\xi}\label{aRind}%
\end{equation}
This constant proper acceleration of the spatial point relative to the
geodesic coordinates of a fixed Minkowski frame gives the relation between the
coordinates of the noninertial frame relative to geodesic coordinates
as\cite{Schutz150}
\begin{equation}
ct=\xi\sinh\eta,\text{ \ }x=\xi\cosh\eta,\label{Rind}%
\end{equation}
while the $y$ and $z$ coordinates are unchanged between the Minkowski and
Rindler coordinate frames. \ 

Since we are dealing here with a scalar radiation field $\phi(ct,x,y,z),$ the
field $\varphi(\eta,\xi,y,z)$ in the Rindler frame takes the same value at the
corresponding point of the Rindler frame,%
\begin{equation}
\phi(ct,x,y,z)=\phi(\xi\sinh\eta,\xi\cosh\eta,y,z)=\varphi(\eta,\xi
,y,z).\label{fRind}%
\end{equation}
Therefore the correlation function for the zero-point field $\varphi_{0}$ in
the Rindler frame follows from Eq. (\ref{zcorr}) as
\begin{align}
&  \left\langle \varphi_{0}(\eta,\xi,y,z)\varphi_{0}(\eta^{\prime},\xi
^{\prime},y^{\prime},z^{\prime})\right\rangle \nonumber\\
&  =\left\langle \phi_{0}(\xi\sinh\eta,\xi\cosh\eta,y,z)\phi_{0}(\xi^{\prime
}\sinh\eta^{\prime},\xi^{\prime}\cosh\eta^{\prime},y^{\prime},z)\right\rangle
\nonumber\\
&  =\frac{-(\hbar c/\pi)}{(\xi\sinh\eta-\xi\sinh\eta)^{2}-(\xi\cosh\eta
-\xi\cosh\eta)^{2}-(y-y^{\prime})^{2}-(z-z^{\prime})^{2}}\nonumber\\
&  =\frac{-(\hbar c/\pi)}{2\xi\xi^{\prime}\cosh(\eta-\eta^{\prime})-\xi
^{2}-\xi^{^{\prime}2}-(y-y^{\prime})^{2}-(z-z^{\prime})^{2}}.\label{Rindcorr}%
\end{align}

The zero-point correlation function (\ref{Rindcorr}) still depends upon the
geodesic separation between the spacetime points at which it is evaluated, but
now the coordinates used in the evaluation are those of the noninertial
Rindler frame. \ The Rindler zero-point correlation function in Eq.
(\ref{Rindcorr}) at a single time $\eta=\eta^{\prime}$ takes exactly the same
form $|\mathbf{r-r}^{\prime}|^{-2}$ (involving the spatial separation) as the
Minkowski zero-point correlation function in Eq. (\ref{zcorr}) at the single
time $t=t^{\prime}$, because a single time in the Rindler frame is also a
single time in the Minkowski frame which is instantaneously at rest with
respect to the coordinates of the Rindler frame, and the Rindler spatial
coordinates can be used as geodesic coordinates in the associated Minkowski
frame. \ 

It is the concept of time which is so vastly different between the Rindler
frame (or any noninertial frame) and any Minkowski frame. The correlation
function for the zero-point field $\varphi_{0}$ at a single spatial point
$\xi,y,z$ in the Rindler frame but at two different times $\eta,\eta^{\prime}$
follows from Eq. (\ref{Rindcorr}) and gives%
\begin{align}
\left\langle \varphi_{0}(\eta,\xi,y,z)\varphi_{0}(\eta^{\prime},\xi
,y,z)\right\rangle  &  =\frac{-(\hbar c/\pi)}{2\xi^{2}\cosh(\eta-\eta^{\prime
})-2\xi^{2}}\nonumber\\
&  =\frac{-(\hbar c/\pi)}{4\xi^{2}\sinh^{2}[(\eta-\eta^{\prime})/2]}%
\nonumber\\
&  =\frac{-(\hbar c/\pi)}{4\xi^{2}\sinh^{2}[c(\tau-\tau^{\prime})/(2\xi
)]}\label{ttRind}%
\end{align}
where in the last line of Eq. (\ref{ttRind}) we have introduced the local
proper time $\tau=\xi\eta/c$ at the spatial coordinates $\xi,y,z.$ \ We notice
that for small time intervals, $\eta-\eta^{\prime}<<1,$ this correlation
function takes the form $\left\langle \varphi_{0}(\eta,\xi,y,z)\varphi
_{0}(\eta^{\prime},\xi,y,z)\right\rangle =-(\hbar c/\pi)/[c^{2}(\tau
-\tau^{\prime})^{2}]$ where $\tau-\tau=\xi(\eta-\eta^{\prime})/c$ is the
proper-time difference between the points. \ This correlation function
expression involving $(\tau-\tau^{\prime})^{-2}$ agrees with the corresponding
expression in Eq. (\ref{zcorr}) involving $(t-t^{\prime})^{-2}$ in the
Minkowski frame. \ However, at large time intervals, $\eta-\eta^{\prime}>>1,$
the functional form in (\ref{ttRind}) changes completely and involves
exponentials of the time interval. \ Thus in this noninertial Rindler frame,
the zero-point correlation function (\ref{Rindcorr}) now has acquired a
feature relative to the time coordinate $\eta$ in the Rindler frame. \ The
time interval $\eta_{tr}=$ $\eta-\eta^{\prime}=1$ can be taken as a transition
time interval for the correlation function when evaluated in the coordinates
of the Rindler frame. \ Associated with this transition time $\eta_{tr}=1$,
there will be a transition frequency $\kappa_{tr}=1/\eta_{tr}=1.$ \ Now when
we apply a time-dilating symmetry transformation to the zero-point fields, the
correlation function (\ref{ttRind}) for the zero-point fields is no longer an
invariant but rather is carried into a new correlation function associated
with a new transition time and a new transition frequency. \ This new
correlation function for $\sigma>1$ corresponds to a situation of thermal
radiation equilibrium at positive empirical temperature in the noninertial
frame. \ 

In a Minkowski frame, a time-dilating symmetry transformation $t\rightarrow
\overline{t}=\sigma t$ corresponds to a uniform scale transformation in space
and time as given above in Eq. (\ref{dil}) with the corresponding scale change
of the metric in Eq. (\ref{mdil}). \ However, in a noninertial frame, a
time-dilating symmetry transformation $\eta\rightarrow\overline{\eta}%
=\sigma\eta$ for the radiation fields involves complicated transformations of
the spatial coordinates; a time-dilating symmetry transformation in a
noninertial frame is a time-dilating \textit{conformal} transformation, where
for Rindler coordinates, the metric is transformed as
\begin{equation}
ds^{2}\rightarrow\,d\overline{s}^{2}=\sigma^{2}[f(\sigma,\xi)]^{2}(\xi
^{2}d\eta^{2}-d\xi^{2}-dy^{2}-dz^{2}).\label{conf}%
\end{equation}
\ In a Rindler frame in \textit{two} spacetime dimensions, the time-dilating
conformal transformation has been given explicitly.\cite{str} \ \ Here, where
we are working in \textit{four} spacetime dimensions, we merely note that a
time-dilating conformal transformation will transform the radiation field
$\varphi$ by the local scaling factor $\mathfrak{s=}\sigma f(\sigma,\xi)$ in
Eq. (\ref{conf}) while the spatial coordinate $\xi$ is transformed as
$\xi\rightarrow\overline{\xi}=f(\sigma,\xi)\xi$ and the time coordinate as
$\eta\rightarrow\overline{\eta}=\sigma\eta$ so that the correlation function
transforms as
\begin{align}
\left\langle \overline{\varphi}_{\sigma}(\eta,\xi,y,z)\overline{\varphi
}_{\sigma}(\eta^{\prime},\xi,y,z)\right\rangle  &  =\frac{-[\sigma
f(\sigma,\xi)]^{2}(\hbar c/\pi)}{4[f(\sigma,\xi)]^{2}\xi^{2}\sinh^{2}%
[\sigma(\eta-\eta^{\prime})/2]}\nonumber\\
&  =\frac{-\sigma^{2}(\hbar c/\pi)}{4\xi^{2}\sinh^{2}[\sigma(\eta-\eta
^{\prime})/2]}\nonumber\\
&  =\frac{-\sigma^{2}(\hbar c/\pi)}{4\ \xi^{2}\sinh^{2}[c\sigma(\tau
-\tau^{\prime})/(2\xi)]}.\label{sRind}%
\end{align}
We notice that for very short times $\eta-\eta^{\prime}<<1/\sigma$, the
correlation function in Eq. (\ref{sRind}) still goes over to the dependence
$\xi^{-2}(\eta-\eta^{\prime})^{-2}=[c(\tau-\tau^{\prime})]^{-2}$ appropriate
for zero-point radiation in the local momentarily comoving reference frame.
However, now this short-time for $\eta-\eta^{\prime}$ must be small compared
to both the transition time $\eta_{tr}=1$ associated with the use of
noninertial Rindler coordinates and also small compared to $1/\sigma$
associated with the thermal contribution to the correlation function.

In a general coordinate frame, the local absolute temperature $T$ in the
region of thermal equilibrium follows the Tolman-Ehrenfest relation\cite{Tolm}
$T(g_{00})^{1/2}=const$. \ In a Minkowski frame where $g_{\text{M}00}=1$ is a
constant, the temperature $T$ is uniform throughout the region. However, in a
noninertial frame, the local temperature $T$ will vary with the spatial
coordinate value. For a Rindler frame, $(g_{00})^{1/2}=\xi,$ so that $T\xi$
equals a global constant throughout the region of thermal equilibrium. For the
situation of our analysis where zero-point radiation in the Rindler frame is
transformed by a time-dilating conformal transformation into thermal radiation
at positive absolute temperature, the scale factor $\sigma$ provides a global
empirical temperature so that the Tolman-Ehrenfest relation takes the form
$T\xi=h(\sigma)$ where $h(\sigma)$ is some function of $\sigma$.

In order to derive the spectrum of thermal radiation in an inertial Minkowski
frame, we must bring the spatial coordinate point (where the thermal
correlation function is evaluated) out to the asymptotic region of the
noninertial frame were the proper acceleration becomes ever smaller. For a
Rindler frame, this means taking the coordinate $\xi$ to spatial infinity
while holding the correlation proper-time difference $\tau-\tau^{\prime}$
fixed. For zero-point radiation in Eq. (\ref{ttRind}), the argument of the
hyperbolic sine function involves $(\tau-\tau^{\prime})/\xi$ and so becomes
ever smaller as $\xi\rightarrow\infty;$ this brings the correlation function
back to the \textit{zero-point} radiation correlation function expression
involving $(\tau-\tau^{\prime})^{-2}$, exactly as in the correlation function
(\ref{zcorr}) appropriate for Minkowski spacetime. In order to derive the
\textit{thermal} spectrum at positive temperature in a Minkowski frame, we
must hold the local temperature fixed at a constant value in Eq. (\ref{sRind})
by increasing $\sigma$ as we increase the coordinate $\xi$ so that $\sigma
/\xi$ is held constant as the coordinate point is taken to the asymptotic
region. \ Once we are in the asymptotic region where the proper acceleration
has become negligible, the only contribution to the correlation function is
due to the thermal radiation spectrum. \ Therefore we simply carry out the
Fourier time transform to obtain the thermal spectrum associated with the
correlation function to find
\begin{equation}
\mathcal{E}_{\sigma}(\omega)=\frac{1}{2}\hbar\omega\coth\left(  \frac{\pi
\xi\omega}{c\sigma}\right)  .\label{Esig}%
\end{equation}

We can check the result in Eq. (\ref{Esig}) by going through the more familiar
calculation of the correlation function of the field when starting from a
known energy per normal mode.\cite{rel} \ Thus the relativistic radiation
field is given as%
\begin{equation}
\phi(ct,x,y,z)=\sum_{n_{x},n_{y},n_{z}}\left(  \frac{8\pi k^{2}\mathcal{E}%
}{L_{x}L_{y}L_{z}}\right)  ^{1/2}\cos[\mathbf{k}\cdot\mathbf{r}-ckt-\theta
(\mathbf{k})]\label{field}%
\end{equation}
where $\mathcal{E}$ is the energy per normal mode, $\theta(\mathbf{k})$ is the
random phase associated with wavevector $\mathbf{k}$, and we are using box
normalization where $\mathbf{k}=\widehat{x}(n_{x}2\pi/L_{x})+\widehat{y}%
(n_{y}2\pi/L_{y})+\widehat{z}(n_{z}2\pi/L_{z})$ so that $n_{x},n_{y},n_{z}$
run over all positive and negative integers. \ When calculating the
correlation function $\left\langle \phi(ct,\mathbf{r)}\phi(ct^{\prime
},\mathbf{r})\right\rangle ,$ we average over the random phases, and then
assume that the box is so large that the sum over the values of $\mathbf{k}$
may be replaced by an integral $d^{3}k.$ \ Next we integrate over the angular
directions associated with $\mathbf{k.}$\cite{rel} \ If we assume that the
energy per normal mode is as given in Eq. (\ref{Esig}), then the needed
integral involves $\int_{0}^{\infty}dx\,x\coth(x/2)\cos(bx)=-\pi^{2}\sec
$h$^{2}(\pi b)$ which can be obtained by taking a derivative with respect to
$b$ of $\int_{0}^{\infty}dx\coth(x/2)\sin(bx)=\pi\coth(\pi b).$ \ This last
integral may be obtained from the singular Fourier sine transform $\int%
_{0}^{\infty}dx\sin(bx)=1/b$ and the standard integral $\int_{0}^{\infty
}dx\sin(bx)/[e^{x}-1]=(\pi/2)\coth(\pi b).$\cite{GR}

The constant $c\sigma/\xi$ appearing in Eq. (\ref{Esig}) can be related to the
temperature $T$ by taking the small-frequency limit $\omega\rightarrow0$ and
using the expansion $\coth x=1/x+x/3-x^{3}/45+...$ to obtain $(1/2)\hbar
c\sigma/(\pi\xi)=k_{B}T.$ \ Therefore the thermal energy spectrum in a
Minkowski frame is the familiar Planck spectrum including zero-point radiation%

\begin{equation}
\mathcal{E}(\omega,T)=\frac{1}{2}\hbar\omega\coth\left(  \frac{\hbar\omega
}{2k_{B}T}\right)  =\frac{\hbar\omega}{\exp[\hbar\omega/(k_{B}T)]-1}+\frac
{1}{2}\hbar\omega.\label{Plan}%
\end{equation}
The Planck spectrum follows directly from the structure of relativistic spacetime.

It is worth emphasizing that the physicists who struggled with the problem of
the blackbody radiation spectrum at the turn of the 20th century were
unfamiliar with two crucial ingredients necessary for a classical derivation
of the Planck spectrum. \ These missing ingredients include the idea of
classical zero-point radiation and the importance of relativity.\cite{any}
\ These aspects are still not understood in the textbooks of the present era
where the possibility of classical zero-point radiation is never mentioned,
and where classical radiation equilibrium is calculated erroneously from
incompatible mixtures of nonrelativistic and relativistic ideas.\cite{ejp}

The work described in this article has mathematical connections to the Unruh
effect.\cite{Cris} \ However, the analysis here is purely within classical
physics and represents totally different physics from the claims of quantum
field theory.

Most physicists are unaware of one of the foundational relations in \ Nature
when they do not realize that within classical physics, the Planck blackbody
spectrum reflects the structure of relativistic spacetime. \

\end{document}